\documentclass{elsart}

\usepackage{natbib}

\usepackage[T1]{fontenc}
\usepackage{amsmath}
\usepackage{amssymb}
\usepackage{graphicx}

\begin{document}

\begin{frontmatter}



\title{\textsl{INTEGRAL} \& \textsl{RXTE} Power Spectra of Cygnus~X-1}


\author[a,b]{K.~Pottschmidt}\ead{Katja.Pottschmidt@obs.unige.ch},
\author[c]{J.~Wilms}, \author[d]{M.A.~Nowak}, \author[e]{S.~Larsson},
\author[f]{A.A.~Zdziarski}, and \author[g]{G.G.~Pooley}
\address[a]{Max-Planck-Institut f\"ur extraterrestrische Physik,
  Postfach 1312, 87548 Garching, Germany} 
\address[b]{INTEGRAL Science Data Centre, Chemin d'\'Ecogia 16, 1290
  Versoix, Switzerland} 
\address[c]{Department of Physics, Univ.~Warwick, Gibbet Hill Road,
  Coventry CV4~7AL, UK}
\address[d]{MIT-CXC, NE80-6077, 77 Massachusetts Ave., Cambridge, MA
  02139, USA} 
\address[e]{Stockholm Observatory, AlbaNova University Center, 10691
  Stockholm, Sweden} 
\address[f]{Nicolaus Copernicus Astronomical Center, Bartycka 18,
00-716 Warszawa, Poland} 
\address[g]{Astrophysics, Cavendish Laboratory, Madingley Road, Cambridge
CB3~0HE, UK 
}

\begin{abstract}
   We evaluate 0.03--20\,Hz power spectra of the bright black hole
   binary Cyg~X-1 obtained from non-deconvolved
   \textsl{INTEGRAL}-ISGRI event data. The ISGRI power spectra are
   compared to contemporary \textsl{RXTE}-PCA ones in the same hard
   X-ray energy band of 15--70\,keV. They agree well in shape. Since
   the ISGRI power spectrum of \mbox{Cyg~X-1} is not background
   corrected it lies about an order of magnitude below the PCA
   values. In 2003 a soft outburst of Cyg~X-1 occurred. From the
   \textsl{RXTE}-ASM and \textsl{Ryle} radio long term lightcurves and
   the \textsl{RXTE} spectra we see a canonical ``hard state --
   intermediate state -- soft state'' evolution. We discuss the
   evolution of the power spectra in the 15--70\,keV range which so
   far is much less well studied than that at softer energies. We
   interpret our results regarding the origin of certain variability
   components.

\end{abstract}


\end{frontmatter}

\section{Introduction \& Data Extraction}\label{sec:data}

\begin{figure}
\includegraphics[width=0.85\linewidth]{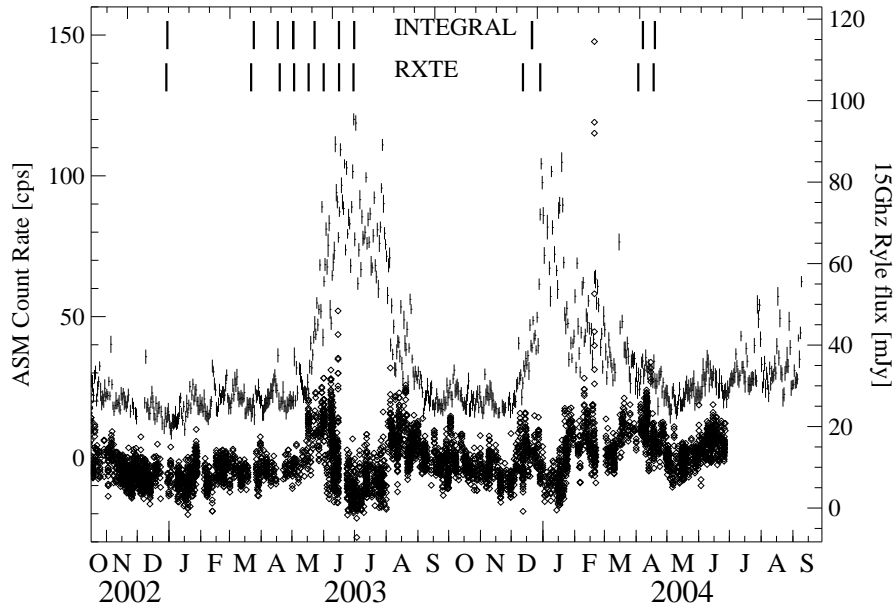}
\caption{\textsl{Ryle} radio flux (15\,GHz) and \textsl{RXTE}-ASM
count rate (2--12\,keV) of Cyg~X-1. The Ryle data points (diamonds)
are plotted without error bars for clarity. Tick-marks denote pointed
\textsl{INTEGRAL} and \textsl{RXTE} observations.}\label{fig:asm}
\end{figure}

Timing studies of black hole binaries allow for a very precise state
classification and are essential for understanding the accretion
process in these objects \citep{kalemci:04,pottschmidt:03}. In this
work we analyze continuum power spectra derived from the Soft
Gamma-Ray Imager \citep[ISGRI;][]{lebrun:03} on-board ESA's
International Gamma-Ray Astrophysics Laboratory
\citep[\textsl{INTEGRAL;}][]{winkler:03}, i.e., of the upper layer of
the coded mask imager IBIS. By comparing contemporaneous
\textsl{RXTE}-PCA and \textsl{INTEGRAL}-ISGRI data of the bright black
hole binary Cyg~X-1 in the 15--70\,keV energy range we show for the
first time that ISGRI can be used to extend black hole timing studies
to the much less well known regime above 20\,keV, where the emission
is dominated by the strongly variable corona/jet system. In the
following we consider all Galactic plane scan (GPS) observations of
\textsl{INTEGRAL} up to 2004 April during which Cyg~X-1 was in the
fully coded field of view of ISGRI (offset$< 4.5^\circ$). Pointed
\textsl{RXTE}-PCA observations from our 14-daily monitoring program
closest in time were chosen for comparison. They are placed into
context of the long term radio and soft X-ray evolution in
Fig.~\ref{fig:asm}: two soft X-ray flaring episodes occurred,
displaying the typical radio quenching associated with black hole soft
states. An analysis of the X-ray spectra confirms this picture
\citep{wilms:04}.

The ISGRI analysis is based on detector events extracted with the
``off-line scientific analysis'' software\footnote{A version
comparable to OSA-4.0 was used. OSA-4.1 qualitatively produces the
same results.} and its event extractor tool ``evts\_extract''. A
selection on ``pixel illumination fraction'' (PIF), i.e., on source
direction, and on energy was performed. Varying the PIF range does not
significantly change the results for this bright source (adopted here:
PIF$\geq$0.8).  Power spectra were calculated following the procedure
described by \citet{pottschmidt:03}. Each ISGRI power spectrum
corresponds to an exposure of typically 1--2\,ks, the PCA power
spectra include 1--6\,ks.

\section{Comparison of ISGRI \& PCA Power Spectra}\label{sec:compare}

\begin{figure}
  \includegraphics[width=0.85\linewidth]{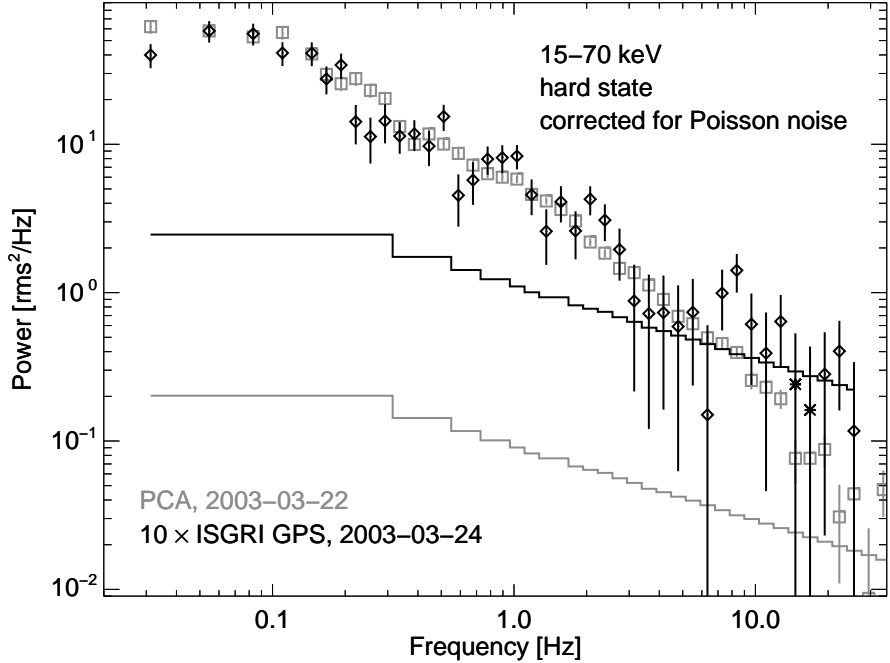}
\caption{\textsl{INTEGRAL}-ISGRI and \textsl{RXTE}-PCA power spectra
of Cyg~X-1 in the 15--70\,keV range, corrected for Poisson noise and
using the normalization of \citet{leahy:83}. In this and the following
figures histograms show the effective noise level corresponding to the
uncertainty of the Poisson noise correction \citep{nowak:99}, stars
denote that the absolute value of a negative power was
plotted.}\label{fig:nonoise}
\end{figure}

Comparing typical PCA (2003--03--22) and ISGRI (2003--03--24) hard
state 15--70\,keV power spectra, we find that they agree well in shape
up to $\sim$5\,Hz where the ISGRI power spectrum becomes noise
dominated (Fig.\ref{fig:nonoise}). However, the variability measured
with ISGRI lies about an order of magnitude too low. This can mainly
be attributed to the much higher ISGRI background and can in principle
be corrected. Note that the normalization of the power spectrum scales
as $1/(S+B)$, where $S$ is the source count rate and $B$ the
background count rate. Part of the offset might also be due to an
energy dependent source rms and the different responses of the two
instruments within the selected band (the photon number weighted
characteristic photon energy of Cyg~X-1 in ISGRI and the PCA is
34--38\,keV and 23--24\,keV, respectively). We also note here that
data from \textsl{INTEGRAL's} performance verification (PV) phase are
generally more difficult to use for this kind of timing analysis:
first, they suffer from many gaps due to telemetry saturation which
have to be taken into account carefully, second, a rise-time event
selection has to be performed for all data obtained before
2003--02--09 (after this date the selection has been performed
on-board). The corrected early power spectra agree reasonably well
with the PCA data, however, they are still noisier than later ISGRI
data (possible reasons: buffer saturation, treatment of noisy pixels,
$\ldots$). The long \mbox{Cyg~X-1} PV phase observations are thus less
well suited for high resolution timing analyses.

\section{Evolution Over the Flare}\label{sec:flare}

\begin{figure}
\includegraphics[width=0.85\linewidth]{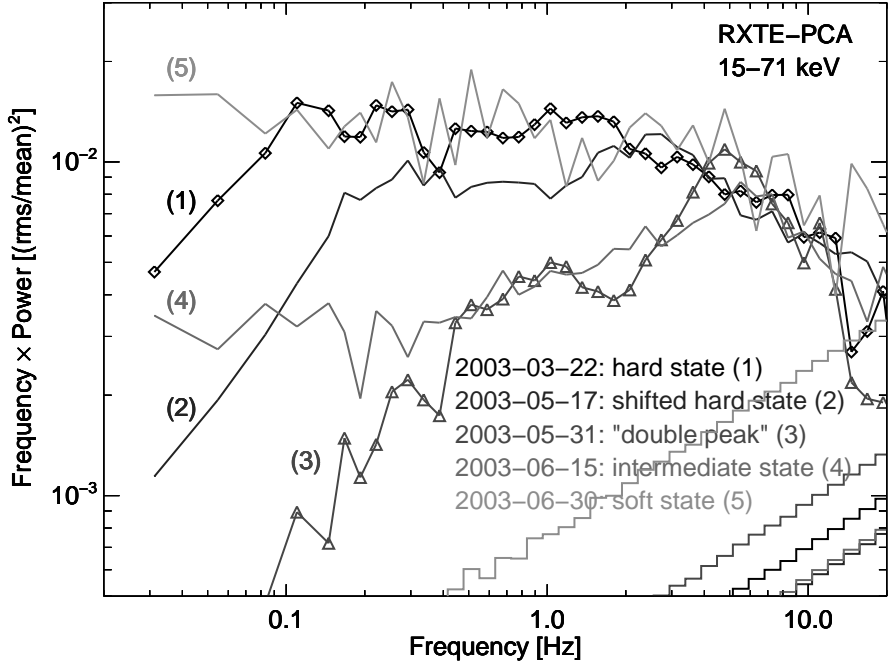} 
\caption{Evolution of the 15--70\,keV \textsl{RXTE}-PCA power spectrum
of Cyg X-1 over the first half of the 2003 flaring episode. For
reasons of clarity the uncertainties of the power values are not
plotted. The hard state and the double peaked power spectrum are
highlighted by diamonds and triangles, respectively. Here and in
Fig.~\ref{fig:ipsdevol} the normalization according to
\citet{miyamoto:92} is used and the power has been multiplied by
frequency.}\label{fig:rpsdevol}
\end{figure}

The rise of the outburst in 2003 is regularly sampled by
\textsl{INTEGRAL} and \textsl{RXTE} observations. We concentrate on
these observations, pointing out that also the remaining ones behave
as expected from Fig.~\ref{fig:asm}. Black hole binary timing behavior
is usually characterized in the \textsl{soft} X-ray band: for the 2003
flare rise we find that the $<4$\,keV PCA power spectra display a
characteristic state transition sequence which is fully consistent
with the spectral evolution. A similar sequence of power spectra has
been observed several times during our 1998--2004 \textsl{RXTE}
monitoring campaign \citep[see, e.g.][]{pottschmidt:03}. In the
15--70\,keV range the changes of the power spectrum follow a similar
pattern (Fig.~\ref{fig:rpsdevol}): Starting from the hard state which
can be described by several broad Lorentzian shaped components, first
the whole shape is shifted toward higher frequencies. A characteristic
double peak appears, developing into the intermediate state dominated
by only one Lorentzian \citep[associated with enhanced time lags and
radio flaring,][]{pottschmidt:03}. Finally the less structured soft
state is reached (power law dominated towards low frequencies). But
there is an important difference to the soft band which to our
knowledge has never been reported before: The ``double peak'' power
spectrum is much more asymmetric at higher energies, further
supporting that the stronger ``enhanced time lag peak'' around
$\sim$4\,Hz in the power spectrum is associated with the hard spectral
component and thus possibly the jet forming region. Also, at these
energies the rms contribution during the soft state is comparable to
that in the hard state, confirming that the variability is primarily
due to the hard spectral component \citep{churazov:01}.

\begin{figure}
  \includegraphics[width=0.65\linewidth]{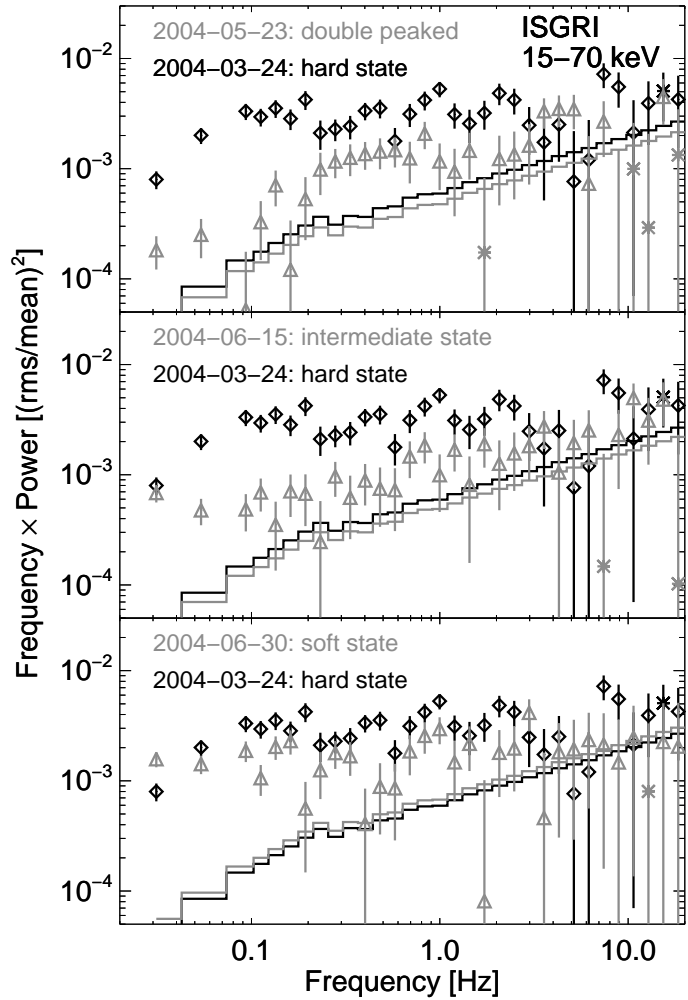}
\caption{Evolution of the \textsl{INTEGRAL}-ISGRI power spectrum of
Cyg~X-1 over the first half of the 2003 flaring episode for energies
between 15 and 70\,keV. In each panel a different evolution state is
compared to the hard state.}\label{fig:ipsdevol}
\end{figure}

ISGRI observations within the same day or within a few days of the PCA
measurements of Fig.~\ref{fig:rpsdevol} are available. The
corresponding 15--70\,keV ISGRI power spectra
(Fig.~\ref{fig:ipsdevol}) follow the same general sequence as the PCA
power spectra, the characteristic differences between the different
states being especially apparent at frequencies below 0.1\,Hz: the
hard state variability is the strongest with the soft state one almost
at the same level. Before the intermediate state is reached the double
peak form is roughly visible. Also consistent with the PCA the
intermediate state is less variable, too close to the effective noise
level to see the ``enhanced time lag peak''. However, the expected
power at low frequencies is apparent in the intermediate state and
also in the soft state. Unfortunately, averaging the ISGRI power
spectra over more than one pointing for clearer results is not
possible with this limited but dynamic data set. The same is true for
power spectra above 70\,keV or in narrower energy bands. A
multi-wavelength observation campaign of Cyg~X-1 is planned for fall
2004 and with 320\,ks of \textsl{INTEGRAL} data should provide this
opportunity, however.

\section{Summary \& Conclusions}\label{sec:con}

\begin{itemize}

\item Contemporary \textsl{INTEGRAL}-ISGRI and \textsl{RXTE}-PCA
15--70\,keV power spectra of Cyg~X-1 are consistent in shape (data up
to 2003--02--09 have to be rise-time filtered and suffer from
additional instrumental effects).
  
\item Before deriving absolute rms values from the ISGRI power
spectra, a background correction is required. In the case of Cyg~X-1
the uncorrected power is an order of magnitude too low. A better
characterization of the background influence is work in progress.
 
\item This is one of only a few studies effectively accessing energies
above the \textsl{RXTE}-PCA range. The fact that the same power
spectral shape is observed for both instruments means that it does not
change much between 20 and 40\,keV and is clearly attributed to
coronal emission. Even higher energies will be accessible with the
upcoming \textsl{INTEGRAL} observations and maybe also with a careful
analysis of the PV phase data.
  
\item The peak at 4--8\,Hz in the power spectrum dominates the
transitional states at all PCA energies while other components only
play a role in the soft band. The additional association with enhanced
time lags and radio flaring speaks for an origin in or near the jet
forming region.

\end{itemize}

\section*{Acknowledgments}
  This work has been financed by Deutsches Zentrum f\"ur Luft- und
  Raumfahrt grants 50~OG~95030, 50~OG~9601, and 50~OG~302 as well as
  the KBN grants PBZ-KBN-054/P03/2001, and 1P03D01827. We thank all
  people involved in building and calibrating \textsl{INTEGRAL} for
  their efforts, and E.~Smith and J.~Swank for the very smooth
  scheduling of the \textsl{RXTE} observations.




\bibliographystyle{elsart-harv}

\end{document}